\documentclass[titlepage]{revtex4-2}

\usepackage{graphicx}% Include figure files
\usepackage{dcolumn}% Align table columns on decimal point
\usepackage{bm}% bold math
\usepackage{amsmath,amsfonts,amssymb}

\def\bx{{\bf x}}
\def\bxs{{\bf x}_S}

\def\bxf{{\bf x}_F}

\def\bxh{{\bf x}_H}
\def\bxhs{{\bf x}_{H,S}}
\def\bxhf{{\bf x}_{H,F}}

\def\bsh{{\bf s}_{H}}

\def\So{{\mathbb{S}_0}}

\def\dt{{\partial_t}}
\def\di{{\partial_i}}

\newcommand{\eqnref}[1]{Equation \eqref{#1}}
\newcommand{\eqnsref}[1]{Equations \eqref{#1}}
\newcommand{\figref}[1]{Figure \ref{#1}}
\newcommand{\figsref}[1]{Figures \ref{#1}}

\graphicspath{{/home/joeri/Documents/Python/1D/Evanescent/paperimage/}{pdf/}}

\begin{document}

\title{On evanescent wave field retrieval with the Marchenko method in 2D settings}
\author{Joeri Brackenhoff}
\affiliation{Quantairra Research and Development B.V.}
\email{jabrackenhoff@quantairra.nl}
\author{Kees Wapenaar}
\affiliation{Delft University of Technology}

\begin{abstract}
We show the capability of the Marchenko method to retrieve not only propagating waves, but also evanescent waves, based on a recent derivation of the Marchenko method that does not depend on up-down decomposition inside the medium of interest. We show how these wave fields can be easily retrieved in the slowness-intercept-time domain and what the wave fields look like when they are transformed back to the space-time domain. It is vital for the retrieval of the coda of the wave field that the initial estimate of the focusing function is a direct arrival that contains both the up-going and down-going component of the evanescent wave field. This is because these events directly overlay each other in time.
\end{abstract}

\maketitle

\newpage
\section*{Theory}
We briefly consider the required theory of the Marchenko method with and without up-down decomposition and the scheme of retrieval.

We assume that we are dealing with an acoustic medium that is defined by compressibility $\kappa(\bx)$ and density $\rho(\bx)$, where $\bx=(x_1,x_2,x_3)^T$ is a location vector in the Cartesian coordinates system, where $x_3$ is pointing downwards. These properties are related to propagation velocity $c(\bx)=(\kappa\rho)^{-1/2}$. Assuming a heterogeneous medium, we can define the linearized equation of motion
%%%%%%%%%%%%%%%%%%%%%%%%%%%%%%%%%%%%%%%%%%%%%%%%%%%%%%%%%%%%%%%%%%%%%%%%%%%%%%%%%%%%%%%%%%%%%%%%%%%%%%%%%%%%%%%%%%%%%%%%%%%%%
\begin{equation} \label{eqmot}
\rho(\bx)\dt v_i(\bx,t) + \di p(\bx,t)=0
\end{equation}
%%%%%%%%%%%%%%%%%%%%%%%%%%%%%%%%%%%%%%%%%%%%%%%%%%%%%%%%%%%%%%%%%%%%%%%%%%%%%%%%%%%%%%%%%%%%%%%%%%%%%%%%%%%%%%%%%%%%%%%%%%%%%
and the linearized equation of deformation
%%%%%%%%%%%%%%%%%%%%%%%%%%%%%%%%%%%%%%%%%%%%%%%%%%%%%%%%%%%%%%%%%%%%%%%%%%%%%%%%%%%%%%%%%%%%%%%%%%%%%%%%%%%%%%%%%%%%%%%%%%%%%
\begin{equation} \label{eqdef}
\kappa\dt p(\bx,t) + \di v_i(\bx,t) = q(\bx,t),
\end{equation}
%%%%%%%%%%%%%%%%%%%%%%%%%%%%%%%%%%%%%%%%%%%%%%%%%%%%%%%%%%%%%%%%%%%%%%%%%%%%%%%%%%%%%%%%%%%%%%%%%%%%%%%%%%%%%%%%%%%%%%%%%%%%%
where $p(\bx,t)$ is a space ($\bx$) and time ($t$) dependent acoustic pressure wave field, $v_i(\bx,t)$ is the $i^{\rm th}$-component of the particle velocity vector $\mathbf{v}(\bx,t)$ and $q(\bx,t)$ is the volume-injection rate density source term. $\dt=\frac{\partial}{\partial t}$ is the operator to apply a temporal derivative and $\di=\frac{\partial}{\partial x_i}$ is the operator to apply a spatial derivative in the $x_i$-direction. The subscript $i$ follows Einstein's summation convention and takes on values of 1,2 and 3.

We apply $\di \rho(\bx)^{-1}$ to \eqnref{eqmot} and $\dt$ to \eqnref{eqdef} and subtract the latter from the former to obtain
%%%%%%%%%%%%%%%%%%%%%%%%%%%%%%%%%%%%%%%%%%%%%%%%%%%%%%%%%%%%%%%%%%%%%%%%%%%%%%%%%%%%%%%%%%%%%%%%%%%%%%%%%%%%%%%%%%%%%%%%%%%%%
\begin{equation} \label{waveq}
\di(\rho(\bx)^{-1}\di p(\bx,t)) - \kappa(\bx)\partial_t^2p(\bx,t) = -\dt q(\bx,t),
\end{equation}
%%%%%%%%%%%%%%%%%%%%%%%%%%%%%%%%%%%%%%%%%%%%%%%%%%%%%%%%%%%%%%%%%%%%%%%%%%%%%%%%%%%%%%%%%%%%%%%%%%%%%%%%%%%%%%%%%%%%%%%%%%%%%
which is the acoustic wave equation in the space-time domain.

For various reasons, in this paper we will consider data that are in the slowness-intercept-time domain. To obtain these data, we first transform the data to the slowness-frequency domain via the following transform
%%%%%%%%%%%%%%%%%%%%%%%%%%%%%%%%%%%%%%%%%%%%%%%%%%%%%%%%%%%%%%%%%%%%%%%%%%%%%%%%%%%%%%%%%%%%%%%%%%%%%%%%%%%%%%%%%%%%%%%%%%%%%
\begin{equation} \label{tfour}
\tilde{p}(\bsh,x_3,\omega) = \int_{-\infty}^{\infty}\int_{-\infty}^{\infty}p(\bx,t){\rm exp}\{i\omega (t-\bsh\cdot\bxh)\}{\rm d}t{\rm d}\bxh,
\end{equation}
%%%%%%%%%%%%%%%%%%%%%%%%%%%%%%%%%%%%%%%%%%%%%%%%%%%%%%%%%%%%%%%%%%%%%%%%%%%%%%%%%%%%%%%%%%%%%%%%%%%%%%%%%%%%%%%%%%%%%%%%%%%%%
where $\omega$ is the angular frequency, $\bxh=(x_1,x_2)^T$ are the horizontal coordinates, $\bsh=(s_1,s_2)^T$ are the horizontal slownesses in sm$^{-1}$ and $i$ is the imaginary unit. Note that often in \eqnref{tfour}, the horizontal wavenumbers $\mathbf{k}_H$ are used instead of the slownesses. The substitution can be easily made by using $\bsh=\frac{\mathbf{k}_H}{\omega}$. Also note that for \eqnref{tfour}, we assume that this transform is performed in a medium that is laterally invariant. The inverse temporal Fourier transform is defined as
%%%%%%%%%%%%%%%%%%%%%%%%%%%%%%%%%%%%%%%%%%%%%%%%%%%%%%%%%%%%%%%%%%%%%%%%%%%%%%%%%%%%%%%%%%%%%%%%%%%%%%%%%%%%%%%%%%%%%%%%%%%%%
\begin{equation} \label{itfour}
p(\bsh,x_3,\tau) = \frac{1}{\pi}\Re\int_{0}^{\infty}\tilde{p}(\bsh,x_3,\omega){\rm exp}\{-i\omega \tau\}{\rm d}\omega,
\end{equation}
%%%%%%%%%%%%%%%%%%%%%%%%%%%%%%%%%%%%%%%%%%%%%%%%%%%%%%%%%%%%%%%%%%%%%%%%%%%%%%%%%%%%%%%%%%%%%%%%%%%%%%%%%%%%%%%%%%%%%%%%%%%%%
where we used the properties $\tilde{p}(\bsh,x_3,-\omega)=\{\tilde{p}(\bsh,x_3,\omega)\}^*$ and $2\Re\{\tilde{p}(\bsh,x_3,\omega)\}=\tilde{p}(\bsh,x_3,\omega)+\{\tilde{p}(\bsh,x_3,\omega)\}^*$ to simplify the inverse Fourier transform. Note that instead of time $t$ we make use of the intercept time $\tau$ \cite{stoffa1989tau}.

A well-known solution to the wave equation is the Green's function $G=G(\bx,\bxs,t)$, which is the solution of the wave equation at $\bx$ for an impulsive source of volume-injection rate density in the space-time domain at $\bxs$,
%%%%%%%%%%%%%%%%%%%%%%%%%%%%%%%%%%%%%%%%%%%%%%%%%%%%%%%%%%%%%%%%%%%%%%%%%%%%%%%%%%%%%%%%%%%%%%%%%%%%%%%%%%%%%%%%%%%%%%%%%%%%%
\begin{equation} \label{Gwaveq}
\di(\rho(\bx)^{-1}\di G(\bx,\bxs,t)) - \kappa(\bx)\partial_t^2G = -\delta(\bx-\bxs)\dt\delta(t).
\end{equation}
%%%%%%%%%%%%%%%%%%%%%%%%%%%%%%%%%%%%%%%%%%%%%%%%%%%%%%%%%%%%%%%%%%%%%%%%%%%%%%%%%%%%%%%%%%%%%%%%%%%%%%%%%%%%%%%%%%%%%%%%%%%%%
In this paper, we consider not only the full wave field, but also decomposed wave field. It has been shown by various authors \cite{ursin1983review,fokkema1993seismic,aki2002quantitative} that wave fields can be decomposed in their up-going and down-going constituents. This decomposition can be done with normalization of the the decomposed wave fields with respect to the pressure or the power-flux density \cite{wapenaar1998reciprocity}. In this paper, we choose to use pressure-normalization, which means the decomposed wave fields relate to the total wave field as
%%%%%%%%%%%%%%%%%%%%%%%%%%%%%%%%%%%%%%%%%%%%%%%%%%%%%%%%%%%%%%%%%%%%%%%%%%%%%%%%%%%%%%%%%%%%%%%%%%%%%%%%%%%%%%%%%%%%%%%%%%%%%
\begin{equation} \label{decomposition}
p(\bx,t) = p^+(\bx,t)+p^-(\bx,t).
\end{equation}
%%%%%%%%%%%%%%%%%%%%%%%%%%%%%%%%%%%%%%%%%%%%%
In \eqnref{decomposition}, the superscripts $+$ and $-$ refer to down-going and up-going propagation, respectively.

\subsection*{Original Marchenko method}
The original Marchenko method that is used in the field of geophysics makes use of the relations between the decomposed Green's functions and decomposed focusing functions. The Green's function cannot only be decomposed in it's propagation direction, but also the radiation direction of the source and therefore consists of four constituents rather than two:
%%%%%%%%%%%%%%%%%%%%%%%%%%%%%%%%%%%%%%%%%%%%%%%%%%%%%%%%%%%%%%%%%%%%%%%%%%%%%%%%%%%%%%%%%%%%%%%%%%%%%%%%%%%%%%%%%%%%%%%%%%%%%
\begin{equation} \label{Gdecomp}
G(\bx,\bxs,t) = G^{+,+}(\bx,\bxs,t)+G^{-,+}(\bx,\bxs,t)+G^{+,-}(\bx,\bxs,t)+G^{-,-}(\bx,\bxs,t),
\end{equation}
%%%%%%%%%%%%%%%%%%%%%%%%%%%%%%%%%%%%%%%%%%%%%
where the first and second superscript denote the propagation direction at the receiver and source, respectively. 

The focusing function $f_1(\bx,\bxf,t)$ is a wave field at $\bx$ that is designed to focus at $t=0$ at the so-called focal location $\bxf$ in the subsurface of the Earth \cite{van2015green}. It can also be decomposed into it's down-going and up-going constituents ( $f^+_1(\bx,\bxf,t)$ and $f^-_1(\bx,\bxf,t)$, respectively) according to \eqnref{decomposition}. The focusing condition of the function is given as \cite{wapenaar2014marchenko}
%%%%%%%%%%%%%%%%%%%%%%%%%%%%%%%%%%%%%%%%%%%%%%%%%%%%%%%%%%%%%%%%%%%%%%%%%%%%%%%%%%%%%%%%%%%%%%%%%%%%%%%%%%%%%%%%%%%%%%%%%%%%%
\begin{equation} \label{f1pcond}
\partial_3f_1^+(\bx,\bxf,t)|_{x_3=x_{3,F}}=-\frac{1}{2}\rho(\bxf)\delta(\bxh-\bxhf)\dt\delta(t),
\end{equation}
%%%%%%%%%%%%%%%%%%%%%%%%%%%%%%%%%%%%%%%%%%%%%
where $\bxh=(x_1,x_2)^T$. Note that the condition in \eqnref{f1pcond} focuses the wave field inside the subsurface, away from the surface of the Earth.  The derivation of the relation between the decomposed Green's functions and focusing functions can be found in various papers \cite{wapenaar2014marchenko,van2015green}, here we only present the relations themselves
%%%%%%%%%%%%%%%%%%%%%%%%%%%%%%%%%%%%%%%%%%%%%%%%%%%%%%%%%%%%%%%%%%%%%%%%%%%%%%%%%%%%%%%%%%%%%%%%%%%%%%%%%%%%%%%%%%%%%%%%%%%%%
\begin{equation} \label{RD2U}
f_1^-(\bxs,\bxf,t) + G^{-,+}(\bxf,\bxs,t) = \int_{\So}\int^{t}_{-\infty}R^{\cup}(\bxs,\bx,t-t')f_1^+(\bx,\bxf,t'){\rm d}t'{\rm d}\bx,
\end{equation}
%%%%%%%%%%%%%%%%%%%%%%%%%%%%%%%%%%%%%%%%%%%%%
\begin{equation} \label{RU2D}
f_1^+(\bxs,\bxf,-t) - G^{+,+}(\bxf,\bxs,t) = \int_{\So}\int^{t}_{-\infty}R^{\cup}(\bxs,\bx,t-t')f_1^-(\bx,\bxf,-t'){\rm d}t'{\rm d}\bx,
\end{equation}
%%%%%%%%%%%%%%%%%%%%%%%%%%%%%%%%%%%%%%%%%%%%%
where $\So$ represent an acquisition surface at the surface of the Earth, which is assumed to be transparent and homogeneous above this surface. We assume that a reflection response $R^{\cup}(\bxs,\bx,t)$ was recorded at this surface, which measured all the up-going waves caused be a downward radiating source. This suggests that the reflection response is simply $G^{-,+}(\bx,\bxs,t)$, however, in the case of pressure-normalization, the reflection response is defined as
%%%%%%%%%%%%%%%%%%%%%%%%%%%%%%%%%%%%%%%%%%%%%
\begin{equation} \label{Rdef}
\partial_3G^{-,+}(\bx,\bxs,t)|_{x_3=x_{3,S}}=\frac{1}{2}\rho(\bx)\dt R^{\cup}(\bxs,\bx,t),
\end{equation}
%%%%%%%%%%%%%%%%%%%%%%%%%%%%%%%%%%%%%%%%%%%%%
which essentially means that the reflection response has a dipole source, rather than a monopole source.

Note that we can use reflection data to relate the up-going focusing function directly to the down-going focusing function 
%%%%%%%%%%%%%%%%%%%%%%%%%%%%%%%%%%%%%%%%%%%%%%%%%%%%%%%%%%%%%%%%%%%%%%%%%%%%%%%%%%%%%%%%%%%%%%%%%%%%%%%%%%%%%%%%%%%%%%%%%%%%%
\begin{equation} \label{f1mcond}
f_1^-(\bxs,\bxf,t) = \int_{\So}\int^{t}_{-\infty}R_T^{\cup}(\bxs,\bx,t-t')f_1^+(\bx,\bxf,t'){\rm d}t'{\rm d}\bx,
\end{equation}
%%%%%%%%%%%%%%%%%%%%%%%%%%%%%%%%%%%%%%%%%%%%%
where $R_T^{\cup}(\bxs,\bx,t)$ is the reflection response of a medium that is truncated below $x_{3,F}$. However, in practice, this type of reflection data is not available, so instead, \eqnsref{RD2U} and \eqref{RU2D} are used to obtain the focusing functions. Because both the focusing functions and Green's functions are unknown and there are only two equations, the system is under-determined. It has been shown that the Green's function and the focusing functions can be separated in time through the use of a windowing function. This window removes all of the Green's function and leaves the focusing function intact, with the exception of the direct arrival $f_{1,d}^+(\bxs,\bxf,t)$. This is because the time-reversed direct arrival of the Green's function $G_d^{+,+}(\bxf,\bxs,-t)$ and $f_{1,d}^+(\bxs,\bxf,t)$ overlap in time at time $-t_d(\bxf,\bxs)$. The window is defined as
%%%%%%%%%%%%%%%%%%%%%%%%%%%%%%%%%%%%%%%%%%%%%
\begin{equation} \label{windowdef}
\theta(\bxf,\bxs)=H(t+t_d(\bxf,\bxs)) - H(t-t_d(\bxf,\bxs)),
\end{equation}
%%%%%%%%%%%%%%%%%%%%%%%%%%%%%%%%%%%%%%%%%%%%%
where $H$ is the Heaviside function. Applying \eqnref{windowdef} to \eqnsref{RD2U} and \eqref{RU2D} yields
%%%%%%%%%%%%%%%%%%%%%%%%%%%%%%%%%%%%%%%%%%%%%
\begin{equation} \label{MD2U}
f_1^-(\bxs,\bxf,t) = \theta(\bxf,\bxs)\int_{\So}\int^{t}_{-\infty}R^{\cup}(\bxs,\bx,t-t')f_1^+(\bx,\bxf,t'){\rm d}t'{\rm d}\bx,
\end{equation}
%%%%%%%%%%%%%%%%%%%%%%%%%%%%%%%%%%%%%%%%%%%%%
\begin{equation} \label{MU2D}
f_1^+(\bxs,\bxf,-t) = f_{1,d}^+(\bxs,\bxf,-t) + \theta(\bxf,\bxs)\int_{\So}\int^{t}_{-\infty}R^{\cup}(\bxs,\bx,t-t')f_1^-(\bx,\bxf,-t'){\rm d}t'{\rm d}\bx.
\end{equation}
%%%%%%%%%%%%%%%%%%%%%%%%%%%%%%%%%%%%%%%%%%%%%
These are the decomposed Marchenko equations and can be solved iteratively. The only requirement is a first estimation of one of the focusing functions, which is usually $f_{1,d}^+(\bxs,\bxf,t)$.

We assume that we are dealing with media that are laterally invariant, so we can consider the wave fields in the slowness-intercept-time domain. We apply \eqnsref{tfour} and \eqref{itfour} to \eqnsref{RD2U} and \eqref{RU2D}
%%%%%%%%%%%%%%%%%%%%%%%%%%%%%%%%%%%%%%%%%%%%%
\begin{equation} \label{RD2Utp}
\tilde{f}_1^-(\bsh,x_{3,S},x_{3,F},\tau) + \tilde{G}^{-,+}(\bsh,x_{3,F},x_{3,S},\tau) = \int^{\tau}_{-\infty}\tilde{R}^{\cup}(\bsh,x_{3,S},x_{3,S},\tau-\tau')\tilde{f}_1^+(\bsh,x_{3,S},x_{3,F},\tau'){\rm d}\tau',
\end{equation}
%%%%%%%%%%%%%%%%%%%%%%%%%%%%%%%%%%%%%%%%%%%%%
\begin{equation} \label{RU2Dtp}
\tilde{f}_1^+(\bsh,x_{3,S},x_{3,F},-\tau) - G^{+,+}(\bsh,x_{3,F},x_{3,S},\tau) = \int^{\tau}_{-\infty}\tilde{R}^{\cup}(\bsh,x_{3,S},x_{3,S},\tau-\tau')\tilde{f}_1^-(\bsh,x_{3,S},x_{3,F},-\tau'){\rm d}\tau',
\end{equation}
%%%%%%%%%%%%%%%%%%%%%%%%%%%%%%%%%%%%%%%%%%%%%
and similarly to \eqnsref{MD2U} and \eqref{MU2D}
%%%%%%%%%%%%%%%%%%%%%%%%%%%%%%%%%%%%%%%%%%%%%
\begin{equation} \label{MD2Utp}
\tilde{f}_1^-(\bsh,x_{3,S},x_{3,F},\tau) = \tilde{\theta}(\bsh,x_{3,F},x_{3,S})\int^{\tau}_{-\infty}\tilde{R}^{\cup}(\bsh,x_{3,S},x_{3,S},\tau-\tau')\tilde{f}_1^+(\bsh,x_{3,S},x_{3,F},\tau'){\rm d}\tau',
\end{equation}
%%%%%%%%%%%%%%%%%%%%%%%%%%%%%%%%%%%%%%%%%%%%%
\begin{equation} \label{MU2Dtp}
\begin{split}
\tilde{f}_1^+(\bsh,x_{3,S},x_{3,F},-\tau) = &\tilde{f}_{1,d}^+(\bsh,x_{3,S},x_{3,F},-\tau) + \\ 
\tilde{\theta}(\bsh,x_{3,F},x_{3,S})\int^{\tau}_{-\infty}&\tilde{R}^{\cup}(\bsh,x_{3,S},x_{3,S},\tau-\tau')\tilde{f}_1^-(\bsh,x_{3,S},x_{3,F},-\tau'){\rm d}\tau'.
\end{split}
\end{equation}
%%%%%%%%%%%%%%%%%%%%%%%%%%%%%%%%%%%%%%%%%%%%%
One advantage of the above equations is that the integral over the horizontal positions has vanished, which allows us to consider the value for each horizontal slowness value separately. These equations can be used to solve the Marchenko method for propagating waves, however, the method fails when evanescent waves are used. This is because in the derivation of \eqnref{RU2D}, a correlation-type reciprocity theorem for decomposed waves is used, which assumes that evanescent waves are ignored \cite{bojarski1983generalized}. \eqnref{RD2U} does not have this issue, as it was derived using a convolution-type reciprocity theorem.

\subsection*{Evanescent Marchenko method}
To overcome the limitation of the Marchenko method for evanescent waves, a recent paper \cite{wapenaar2020evanescent} considered an alternative derivation, that correctly accounted for the evanescent waves. The author considered the wave fields in the slowness-intercept time domain and found that \eqnsref{RD2Utp} and \eqref{RU2Dtp} were valid only when $\bsh\cdot\bsh \leq \frac{1}{c(x_{3,F})^2}$, which is when the wave field is propagating at the focal depth. When the wave field was evanescent at the focal depth but propagating at the surface (i.e. $\bsh\cdot\bsh > \frac{1}{c(x_{3,F})^2}$), only a single equation could be retrieved:
%%%%%%%%%%%%%%%%%%%%%%%%%%%%%%%%%%%%%%%%%%%%%
\begin{equation} \label{ERtp}
\tilde{G}^{-,+}(\bsh,x_{3,F},x_{3,S},\tau) - \tilde{f}_1^+(\bsh,x_{3,S},x_{3,F},-\tau) = \int^{\tau}_{-\infty}\tilde{R}^{\cup}(\bsh,x_{3,S},x_{3,S},\tau-\tau')\tilde{f}_1^+(\bsh,x_{3,S},x_{3,F},\tau'){\rm d}\tau'.
\end{equation}
%%%%%%%%%%%%%%%%%%%%%%%%%%%%%%%%%%%%%%%%%%%%%
The reason this occurred is because the down-going and up-going focusing function can be directly related to each other when the wave field is evanescent, namely by 
%%%%%%%%%%%%%%%%%%%%%%%%%%%%%%%%%%%%%%%%%%%%%
\begin{equation} \label{EFDU}
\tilde{f}_1^-(\bsh,x_{3,S},x_{3,F},\tau) = -\tilde{f}_1^+(\bsh,x_{3,S},x_{3,F},-\tau).
\end{equation}
%%%%%%%%%%%%%%%%%%%%%%%%%%%%%%%%%%%%%%%%%%%%%
This means that \eqnref{ERtp} is an equation with two unknowns, rather than in \eqnsref{RD2Utp} and \eqref{RU2Dtp}, where there where three unknowns per equation. We apply a new window function $\tilde{w}(\bsh,x_{3,F},x_{3,S})$, which is defined as
%%%%%%%%%%%%%%%%%%%%%%%%%%%%%%%%%%%%%%%%%%%%%
\begin{equation} \label{windowdefpt}
\tilde{w}(\bsh,x_{3,F},x_{3,S})=H(\tau_d(\bsh,x_{3,F},x_{3,S})-\tau).
\end{equation}
%%%%%%%%%%%%%%%%%%%%%%%%%%%%%%%%%%%%%%%%%%%%%
When \eqnref{windowdefpt} is applied to \eqnref{ERtp}, the result is
%%%%%%%%%%%%%%%%%%%%%%%%%%%%%%%%%%%%%%%%%%%%%
\begin{equation} \label{ERtpwindow}
\tilde{f}_1^+(\bsh,x_{3,S},x_{3,F},-\tau) = \tilde{f}_{1,d}^+(\bsh,x_{3,S},x_{3,F},-\tau) - \tilde{w}(\bsh,x_{3,F},x_{3,S})\int^{\tau}_{-\infty}\tilde{R}^{\cup}(\bsh,x_{3,S},x_{3,S},\tau-\tau')\tilde{f}_1^+(\bsh,x_{3,S},x_{3,F},\tau'){\rm d}\tau',
\end{equation}
%%%%%%%%%%%%%%%%%%%%%%%%%%%%%%%%%%%%%%%%%%%%%
which can be solved iteratively, as the reflection data are available as a measurement. When the down-going focusing function is retrieved, \eqnref{ERtp} can be used to retrieved the up-going Green's function. However, using this approach, the down-going Green's function is not retrieved. It has been shown in the same work \cite{wapenaar2020evanescent} that the down-going Green's function is related to the up-going Green's function through
%%%%%%%%%%%%%%%%%%%%%%%%%%%%%%%%%%%%%%%%%%%%%
\begin{equation} \label{EGtp}
\tilde{G}^{+,+}(\bsh,x_{3,F},x_{3,S},\tau) = \frac{\rho(x_{3,S})T^+(\bsh,x_{3,F},x_{3,S},\tau)}{2s_3(\bsh,x_{3,S})} + \int^{\tau}_{-\infty}\tilde{R}^{\cap}(\bsh,x_{3,F},x_{3,F},\tau-\tau')\tilde{G}^{-,+}(\bsh,x_{3,F},x_{3,S},\tau'){\rm d}\tau',
\end{equation}
%%%%%%%%%%%%%%%%%%%%%%%%%%%%%%%%%%%%%%%%%%%%%
where $T^+(\bsh,x_{3,F},x_{3,S},\tau)$ is the transmission response of the truncated medium, $\tilde{R}^{\cap}(\bsh,x_{3,F},x_{3,F},\tau)$ is the reflection response 'from below' of the truncated medium and $s_3$ is the vertical slowness, which is related to the horizontal slownesses by
%%%%%%%%%%%%%%%%%%%%%%%%%%%%%%%%%%%%%%%%%%%%%
\begin{equation} \label{slowvert}
s_3(\bsh,x_{3}) = \begin{cases}
+\sqrt{\frac{1}{c(x_{3})^2}-\bsh\cdot\bsh}\quad\rm{for}\quad\bsh\cdot\bsh\leq\frac{1}{c(x_{3})^2}\\
+i\sqrt{\bsh\cdot\bsh-\frac{1}{c(x_{3})^2}}\quad\rm{for}\quad\bsh\cdot\bsh>\frac{1}{c(x_{3})^2}
\end{cases}.
\end{equation}
%%%%%%%%%%%%%%%%%%%%%%%%%%%%%%%%%%%%%%%%%%%%%

The evanescent Marchenko method shown in \cite{wapenaar2020evanescent} is accurate, however, it has two significant downsides. First of all, \eqnref{EGtp} relies on additional information of the medium, although is should be noted that in the work, some approximations are proposed to reduce this burden. The second limitation is that one needs to have a clear indication when the wave field is propagating and when it is evanescent, so that the correct representation can be used.

\subsection*{Marchenko method without up-down decomposition}
All previous uses of the Marchenko method that we have discussed rely on up-down decomposition in some way or form. The issue with the evanescent wave field finds its origins in the decomposed reciprocity theorem of the correlation-type. A more recent derivation of the Marchenko method made no use of decomposition \cite{wapenaar2021green}. The relation between the Green's function and focusing function in this scenario is
%%%%%%%%%%%%%%%%%%%%%%%%%%%%%%%%%%%%%%%%%%%%%
\begin{equation} \label{RF}
G(\bxf,\bxs,t) - f(\bxf,\bxs,-t) = \int_{\So}\int^{t}_{-\infty}R^{\cup}(\bxs,\bx,t-t')f(\bxf,\bx,t'){\rm d}t'{\rm d}\bx,
\end{equation}
%%%%%%%%%%%%%%%%%%%%%%%%%%%%%%%%%%%%%%%%%%%%%
where a new focusing function $f(\bxf,\bxs,t)$ is used which has the focusing condition
%%%%%%%%%%%%%%%%%%%%%%%%%%%%%%%%%%%%%%%%%%%%%%%%%%%%%%%%%%%%%%%%%%%%%%%%%%%%%%%%%%%%%%%%%%%%%%%%%%%%%%%%%%%%%%%%%%%%%%%%%%%%%
\begin{equation} \label{fcond}
\partial_3f(\bx,\bxs,t)|_{x_3=x_{3,S}}=\frac{1}{2}\rho(\bxs)\delta(\bxh-\bxhs)\dt\delta(t).
\end{equation}
%%%%%%%%%%%%%%%%%%%%%%%%%%%%%%%%%%%%%%%%%%%%%
Note that in \eqnref{fcond}, which holds for both propagating and evanescent waves, the focusing occurs at the surface of the Earth, rather than inside the medium like in \eqnref{f1pcond}.

\eqnref{RF} can be expressed in the slowness-intercept time domain as 
%%%%%%%%%%%%%%%%%%%%%%%%%%%%%%%%%%%%%%%%%%%%%
\begin{equation} \label{RFtp}
\tilde{G}(\bsh,x_{3,F},x_{3,S},\tau) - \tilde{f}(\bsh,x_{3,F},x_{3,S},-\tau) = \int^{\tau}_{-\infty}\tilde{R}^{\cup}(\bsh,x_{3,S},x_{3,S},\tau-\tau')\tilde{f}(\bsh,x_{3,F},x_{3,S},\tau'){\rm d}\tau'.
\end{equation}
%%%%%%%%%%%%%%%%%%%%%%%%%%%%%%%%%%%%%%%%%%%%%
By applying the window from \eqnref{windowdefpt}, we can eliminate the Green's function from the left hand side
%%%%%%%%%%%%%%%%%%%%%%%%%%%%%%%%%%%%%%%%%%%%%
\begin{equation} \label{RFWtp}
\tilde{f}(\bsh,x_{3,F},x_{3,S},-\tau) = \tilde{f}_d(\bsh,x_{3,F},x_{3,S},-\tau) - \tilde{w}(\bsh,x_{3,F},x_{3,S})\int^{\tau}_{-\infty}\tilde{R}^{\cup}(\bsh,x_{3,S},x_{3},\tau-\tau')\tilde{f}(\bsh,x_{3,F},x_{3},\tau'){\rm d}\tau',
\end{equation}
%%%%%%%%%%%%%%%%%%%%%%%%%%%%%%%%%%%%%%%%%%%%%
where $\tilde{f}_d(\bsh,x_{3,F},x_{3,S},\tau)$ is the direct arrival of the full focusing function $\tilde{f}(\bsh,x_{3,F},x_{3,S},\tau)$. \eqnref{RFWtp} can be solved iteratively to obtain the focusing function.

Note that \eqnref{RFWtp} is very similar to \eqnref{ERtpwindow}, however, no decomposition is used in \eqnref{RFWtp}, so evanescent waves are accounted for inside the medium of interest. Unlike \eqnref{ERtpwindow} however, we do not need to consider separate equations for the up-going and down-going wave field depending on whether the wave field is propagating or evanescent.

\newpage
\section*{Numerical Results}
\begin{figure}[!htpb]
\centering
\includegraphics[width=0.85\textwidth]{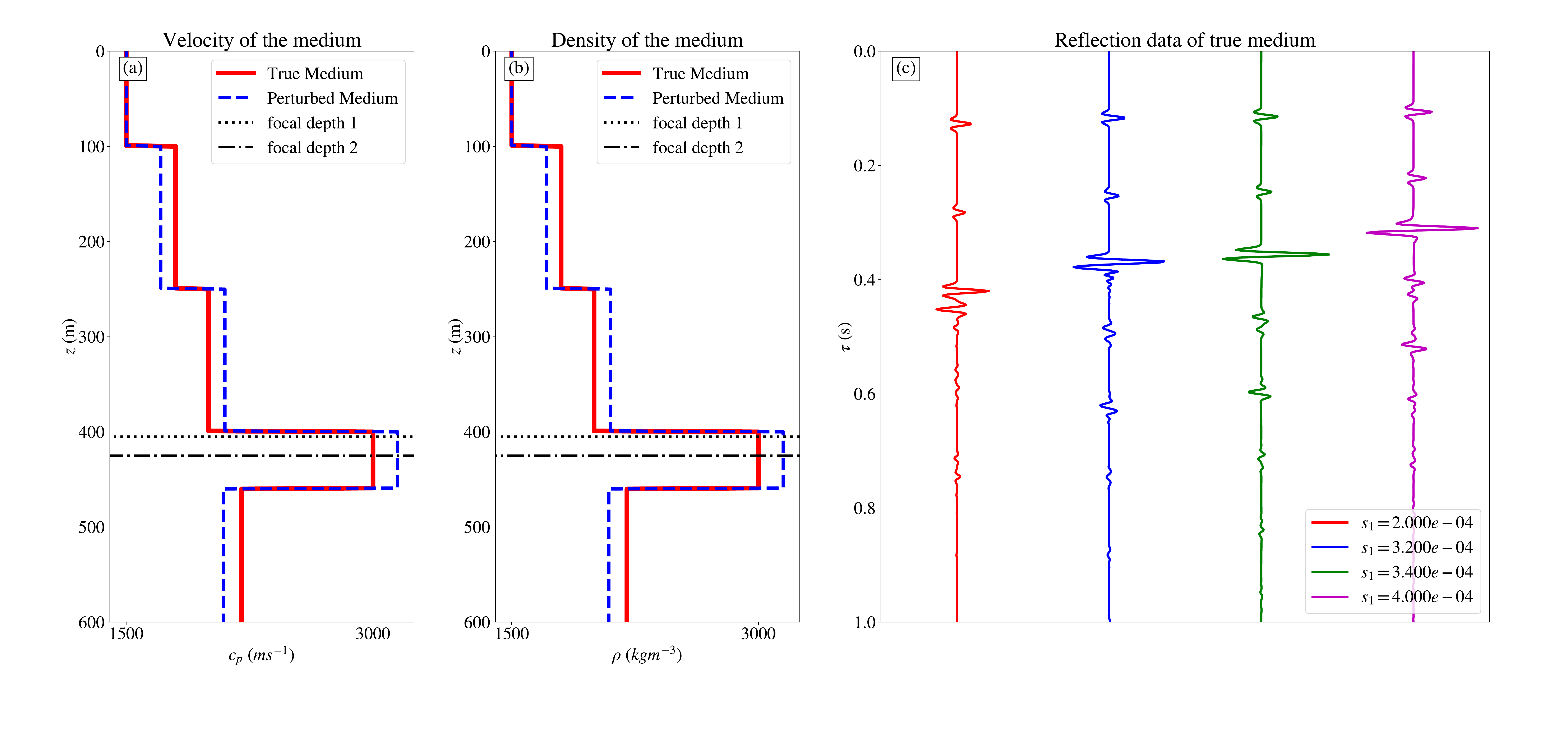}
\caption{Subsurface model for (a) P-wave velocity (ms$^{-1}$) and (b) density (kgm$^{-3}$). The solid red line indicates the full true medium and the dashed blue line indicates a perturbed medium. (c) Reflection data simulated at the top of the true medium as indicated by the solid red lines in (a) and (b). The wave fields in (c) have been convolved with a 50Hz Ricker wavelet and are displayed in the rayparameter-intercept time domain. The various colors indicate different slowness values for the wave fields.}
\label{fig:model}
\end{figure}

To demonstrate that we can retrieve the evanescent wave field using the Marchenko method, we show the application on numerical data. The data were generated using a code that generates a reflection response and transmission response for a 1D medium in the slowness-intercept time domain, given a specific slowness value. \figref{fig:model}a-b shows the velocity model and the density model, respectively as the solid red line. The dashed blue line is a perturbed medium, that has the incorrect medium parameters, which we use to show the effect of errors. The black dotted and dashed-dotted lines indicate two focal depths, for which we will retrieve the focusing functions and Green's functions using the Marchenko method. In \figref{fig:model}c, four reflections responses are shown, for different values of the horizontal slowness. The first trace has a slowness value of 0.0002sm$^{-1}$, which for the layer where we have placed our focal depths is well within the regime of propagating waves. The second and third traces are reflections responses that have slowness values that are just before (0.00032sm$^{-1}$) and after (0.00034sm$^{-1}$) the wave field becomes evanescent, respectively, which occurs at the focal depth at 0.000333sm$^{-1}$. The fourth trace is for a slowness value of 0.0004sm$^{-1}$, well within the evanescent regime.
%%%%%%%%%%%%%%%%%%%%%%%%%%%%%%%%%%%%%%%%%%%%%%%%%%%%%%%%%%%%%%%%%%%%%%%%%%%%%%%%%%%%%%%%%%%%%%%%%%%%%%%%%%%%%%%%%%%%%%%%%%%%%
%%%%%%%%%%%%%%%%%%%%%%%%%%%%%%%%%%%%%%%%%%%%%%%%%%%%%%%%%%%%%%%%%%%%%%%%%%%%%%%%%%%%%%%%%%%%%%%%%%%%%%%%%%%%%%%%%%%%%%%%%%%%%

\subsection*{Retrieval comparisons}

To check the accuracy of our method, we start by modeling the exact focusing function in the medium. It has been shown that the down-going component of the focusing function in \eqnsref{RD2Utp} and \eqref{RU2Dtp} can be directly related to the down-going transmission response of the truncated medium \cite{wapenaar2020evanescent}
%%%%%%%%%%%%%%%%%%%%%%%%%%%%%%%%%%%%%%%%%%%%%
\begin{equation} \label{f1pT}
\tilde{f}_1^+(\bsh,x_{3,S},x_{3,F},\omega) = \frac{\rho(x_{3,F})}{2s_3(\bsh,x_{3,F})\tilde{T}^+(\bsh,x_{3,F},x_{3,S},\omega)},
\end{equation}
%%%%%%%%%%%%%%%%%%%%%%%%%%%%%%%%%%%%%%%%%%%%%
which in turn can be related to the up-going component of the focusing function in \eqnref{RFWtp}
%%%%%%%%%%%%%%%%%%%%%%%%%%%%%%%%%%%%%%%%%%%%%
\begin{equation} \label{fmcond}
\tilde{f}^-(\bsh,x_{3,F},x_{3,S},\tau) = \tilde{f}_1^+(\bsh,x_{3,S},x_{3,F},\tau).
\end{equation}
%%%%%%%%%%%%%%%%%%%%%%%%%%%%%%%%%%%%%%%%%%%%%
To obtain the full focusing function, we require the down-going component of the focusing function $f$. This can be obtained by using a relation, similar to \eqnref{f1mcond},
%%%%%%%%%%%%%%%%%%%%%%%%%%%%%%%%%%%%%%%%%%%%%%%%%%%%%%%%%%%%%%%%%%%%%%%%%%%%%%%%%%%%%%%%%%%%%%%%%%%%%%%%%%%%%%%%%%%%%%%%%%%%%
\begin{equation} \label{fpcond}
\tilde{f}^+(\bsh,x_{3,F},x_{3,S},\tau) = \int^{t}_{-\infty}\tilde{R}^{\cap}(\bsh,x_{3,F},x_{3,F},\tau-\tau')\tilde{f}^-(\bsh,x_{3,F},x_{3,S},\tau'){\rm d}t'.
\end{equation}
%%%%%%%%%%%%%%%%%%%%%%%%%%%%%%%%%%%%%%%%%%%%
The reference focusing function $\tilde{f}$ can then be obtained by adding $\tilde{f}^+$ to $\tilde{f}^-$. Note that we used the up-going and down-going component only to model the full focusing function and because these decomposed wave fields were directly modeled, the propagating and evanescent wave fields are all accounted for.
 
\begin{figure}[!htpb]
\centering
\includegraphics[clip,trim={4cm 19cm 4cm 8cm},width=0.85\textwidth]{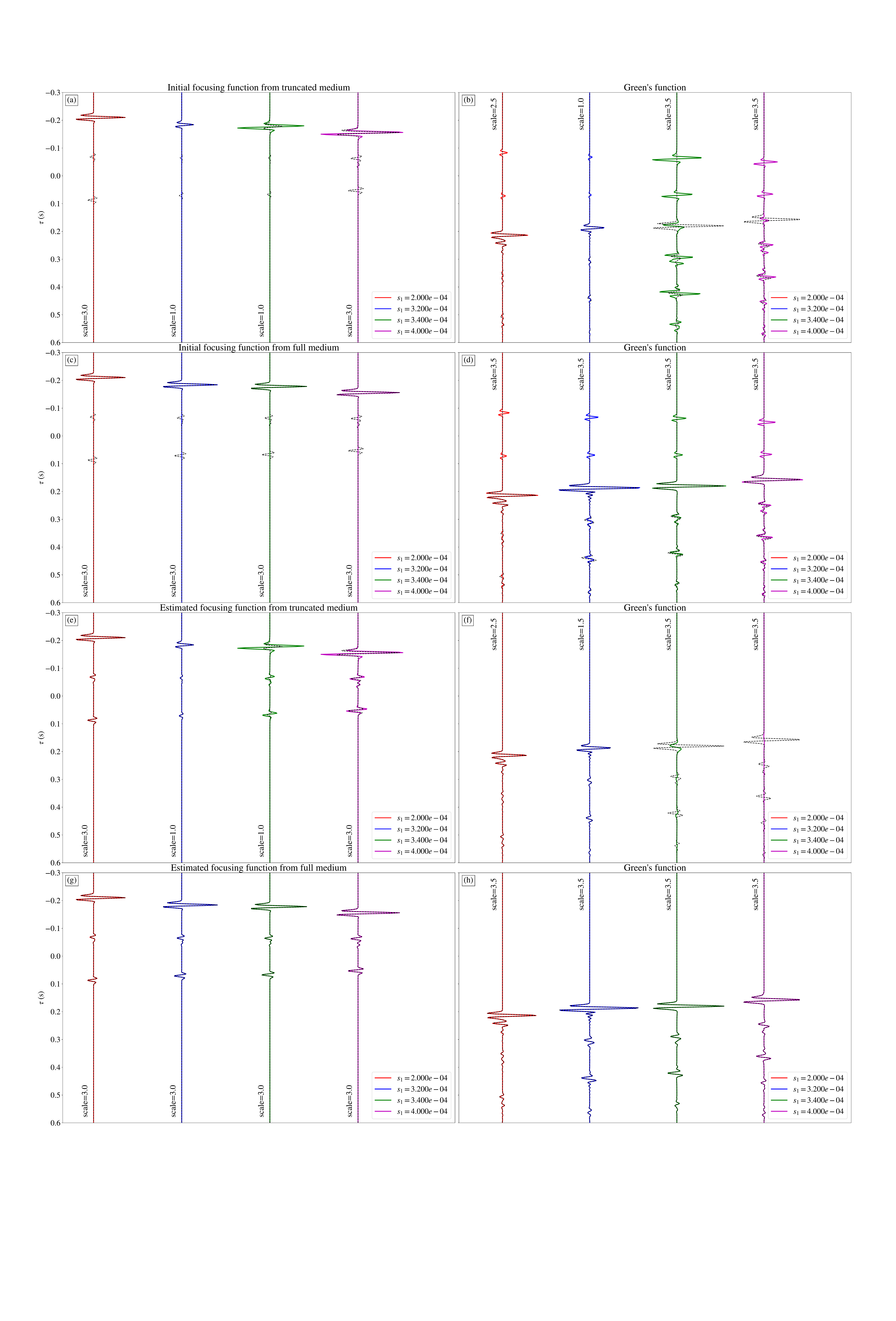}
\caption{wave fields in the subsurface convolved with a 50Hz Ricker wavelet for various ray parameters. The estimated wave fields are shown in solid color, while reference solutions are shown in dotted black. The focal depth is equal to 405m and is indicated by the dotted vertical black line in \figsref{fig:model}a-b. The left column shows focusing functions and the right column shows the corresponding Green's functions. Initial focusing functions estimated in (a) the truncated medium and (c) the full medium. (e) The full focusing function obtained by the Marchenko method using the initial focusing function from (a). (g) Idem as (e) using the initial focusing functions from (c). For the initial estimations of the focusing function, the true medium parameters, as indicated by the solid red line in \figref{fig:model}a-b were used. Note that the reference solutions of the Green's functions and focusing functions were obtained independently from each other. For the purpose of displaying the data, the traces were scaled by various factors, which are indicated next to the traces.}
\label{fig:realmar1}
\end{figure}

We show the Marchenko approach in \figref{fig:realmar1} for the first focal depth that is indicated by the black dotted line in \figref{fig:model}, which is just below the top of the layer of interest. The left column shows focusing functions and the right columns shows the associated Green's functions. The black dashed lines are the reference focusing functions $\tilde{f}$ and the reference Green's functions $\tilde{G}$. The latter were obtained using a direct modeling code, that was performed independently from the reference focusing functions. In the upper row, we show as the colored lines, the input to the Marchenko method, namely the direct arrival $\tilde{f}^-_d$, that we obtained from the truncated medium. We modeled the transmission response in this truncated medium, and used \eqnsref{f1pT} and \eqref{fmcond} to obtain the up-going component of the focusing function and then removed all other events aside from the direct arrival. When this first estimation is used in \eqnref{RFtp}, the result shows that the desired events are not fully recovered and that artifacts are present. We solve \eqnref{RFWtp} with the aim to obtain the full focusing functions and show the result in the third row of the figure. Note that while the first two traces are accurate, when the wave field is propagating, the results for the latter two traces, when the wave field is evanescent, are off, even though we used the same equation for both. While the retrieved focusing functions are somewhat accurate, the Green's functions that are obtained are completely incorrect.

The previous result is puzzling because it was expected that the method would retrieve the evanescent wave field. When \figref{fig:realmar1}a is studied, the cause is revealed however. Note that the initial focusing functions for the propagating wave field match the direct arrival exactly, while this is not the case for the evanescent wave field. This is because an evanescent wave has vertical wave fronts and the up-going and down-going component directly overlay each other in time. When the truncated medium is used to obtain the direct arrival, only the up-going component is employed. To confirm this is indeed the culprit, the full focusing function is modeled by also using \eqnref{fpcond} and the direct arrivals are separated from the coda of the wave field. These initial focusing functions are shown in the left column of the second row of \figref{fig:realmar1}. Note that all direct arrivals now match those of the reference focusing functions. The retrieved focusing functions are shown in the fourth row. Now, the retrieved focusing functions match the reference focusing functions for both propagating and evanescent waves. When the Green's functions in the right column are considered, the result is even more promising. In the fourth row, the retrieved Green's functions match the reference Green's functions in all cases.

%%%%%%%%%%%%%%%%%%%%%%%%%%%%%%%%%%%%%%%%%%%%%%%%%%%%%%%%%%%%%%%%%%%%%%%%%%%%%%%%%%%%%%%%%%%%%%%%%%%%%%%%%%%%%%%%%%%%%%%%%%%%%
%%%%%%%%%%%%%%%%%%%%%%%%%%%%%%%%%%%%%%%%%%%%%%%%%%%%%%%%%%%%%%%%%%%%%%%%%%%%%%%%%%%%%%%%%%%%%%%%%%%%%%%%%%%%%%%%%%%%%%%%%%%%%

\begin{figure}[!htpb]
\centering
\includegraphics[clip,trim={4cm 19cm 4cm 8cm},width=0.85\textwidth]{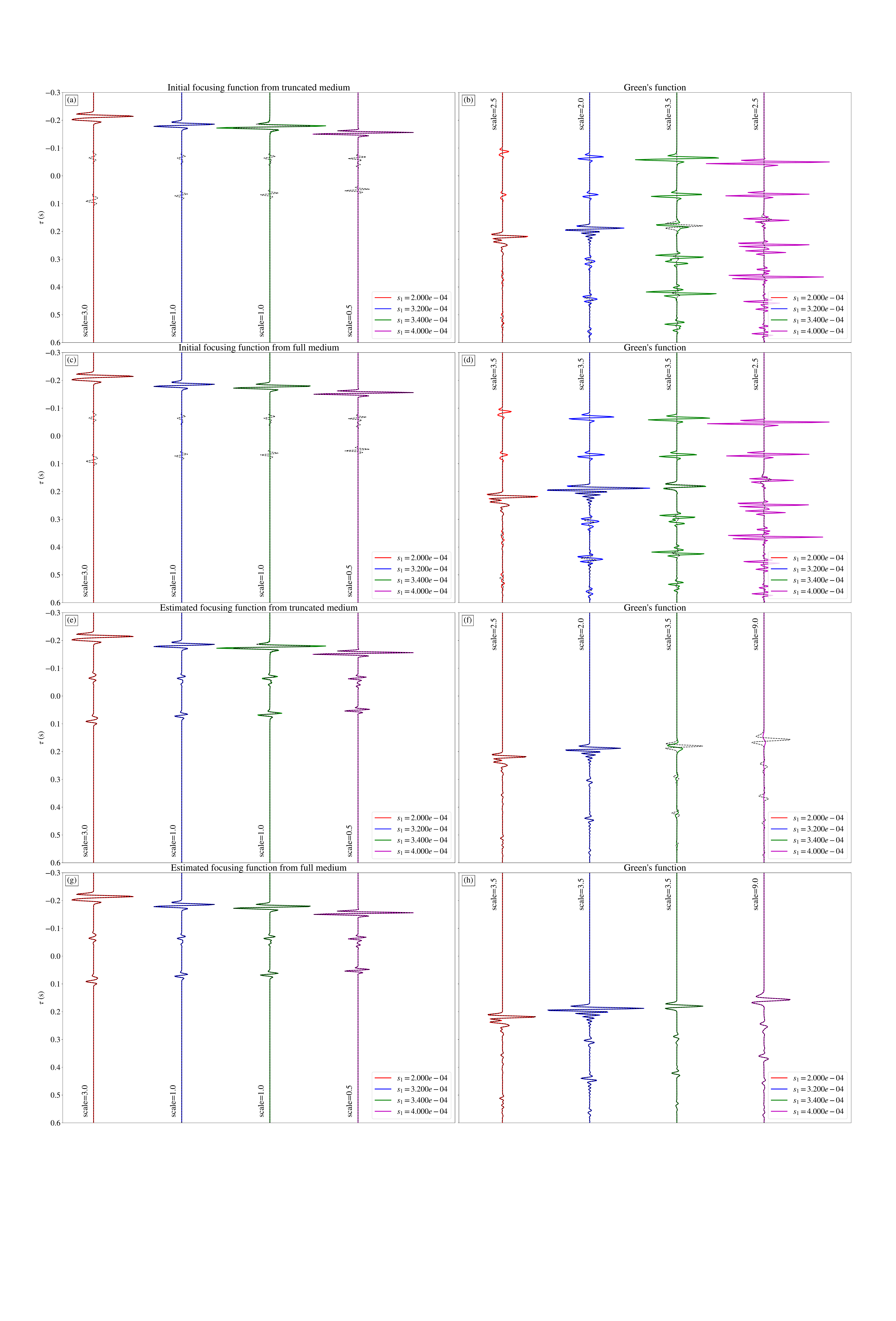}
\caption{Idem as \figref{fig:realmar1}, for a focal depth of 425m, as indicated by the dotted-dashed vertical black line in \figref{fig:model}a-b.}
\label{fig:realmar2}
\end{figure}

To further study the ability of the method to retrieve the evanescent wave field, we consider the second focal depth, indicated by the black dashed-dotted line in \figref{fig:model}, the results of which are shown in \figref{fig:realmar2}. The results are similar to those we found in \figref{fig:realmar1}. The retrieval of the propagating wave field is accurate when the initial focusing functions are estimated using only the truncated medium, but only when the full medium is used to estimate the initial focusing functions, is the evanescent wave field retrieved properly. The evanescent effects are much more pronounced however, because the wave field is estimated deeper inside the medium of interest. This is because the amplitude of the wave field depends exponentially on the depth inside the layer \cite{fokkema1993seismic} and can become unstable, particularly for higher frequencies, quickly. However, the results still support that the coda of the evanescent wave field can be fully retrieved from the direct arrival of the focusing function.

%%%%%%%%%%%%%%%%%%%%%%%%%%%%%%%%%%%%%%%%%%%%%%%%%%%%%%%%%%%%%%%%%%%%%%%%%%%%%%%%%%%%%%%%%%%%%%%%%%%%%%%%%%%%%%%%%%%%%%%%%%%%%
%%%%%%%%%%%%%%%%%%%%%%%%%%%%%%%%%%%%%%%%%%%%%%%%%%%%%%%%%%%%%%%%%%%%%%%%%%%%%%%%%%%%%%%%%%%%%%%%%%%%%%%%%%%%%%%%%%%%%%%%%%%%%

\begin{figure}[!htpb]
\centering
\includegraphics[clip,trim={4cm 19cm 4cm 8cm},width=0.85\textwidth]{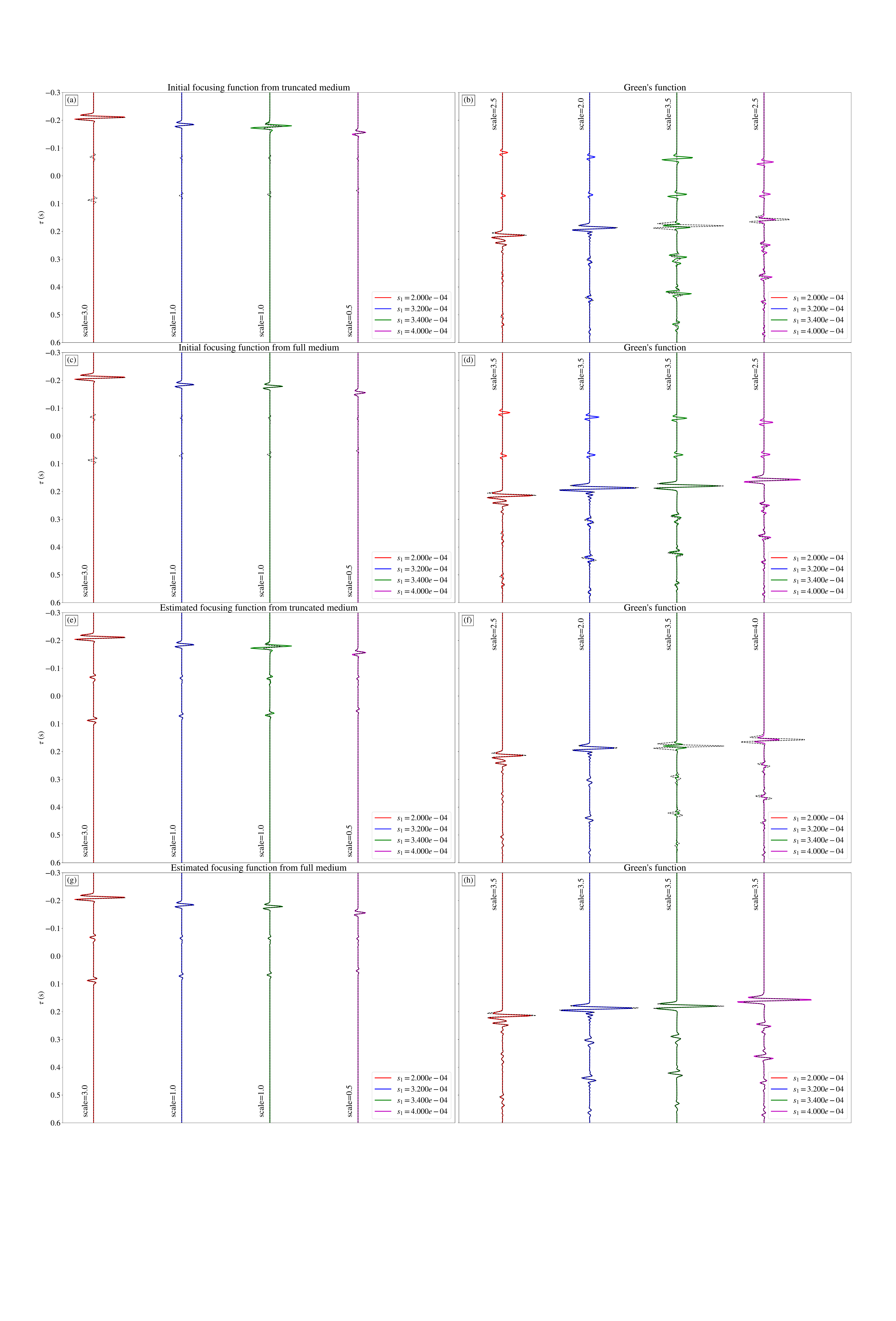}
\caption{Idem as \figref{fig:realmar1}, however, the initial focusing functions in (a) and (c) were estimated using perturbed medium parameters, which are indicated by the dashed blue lines in \figref{fig:model}a-b.}
\label{fig:wrongmar1}
\end{figure}

The sensitivity of the retrieval of the focusing function to the direct arrival does raise a potential issue. When only the full direct arrival can be used, the retrieval in practice will be very difficult. To gauge the sensitivity of the method, we repeat the experiment in \figref{fig:realmar1}, but this time, we use the perturbed medium parameters that are shown by the dashed blue lines in \figref{fig:model}a-b to model the transmission response and reflection response 'from below' that are used in \eqnsref{f1pT}-\eqref{fpcond}, while keeping the reflection response at the surface, which is used in \eqnref{RFWtp}, the same as before. The results of are shown in \figref{fig:wrongmar1}. The results show that while the exact focusing functions and Green's functions cannot be retrieved, the results are still fairly accurate and the result does not immediately degrade, even for the evanescent wave field. The sensitivity to using the direct arrival that was obtained from the full medium is something that cannot be avoided, however, as long as both components of the direct evanescent wave field effect are included in some way, an acceptable result can be obtained.

%%%%%%%%%%%%%%%%%%%%%%%%%%%%%%%%%%%%%%%%%%%%%%%%%%%%%%%%%%%%%%%%%%%%%%%%%%%%%%%%%%%%%%%%%%%%%%%%%%%%%%%%%%%%%%%%%%%%%%%%%%%%%
%%%%%%%%%%%%%%%%%%%%%%%%%%%%%%%%%%%%%%%%%%%%%%%%%%%%%%%%%%%%%%%%%%%%%%%%%%%%%%%%%%%%%%%%%%%%%%%%%%%%%%%%%%%%%%%%%%%%%%%%%%%%%

\subsection*{Results in the space-time domain}

\begin{figure}[!htpb]
\centering
\includegraphics[clip,trim={4cm 5cm 1cm 5cm},width=0.95\textwidth]{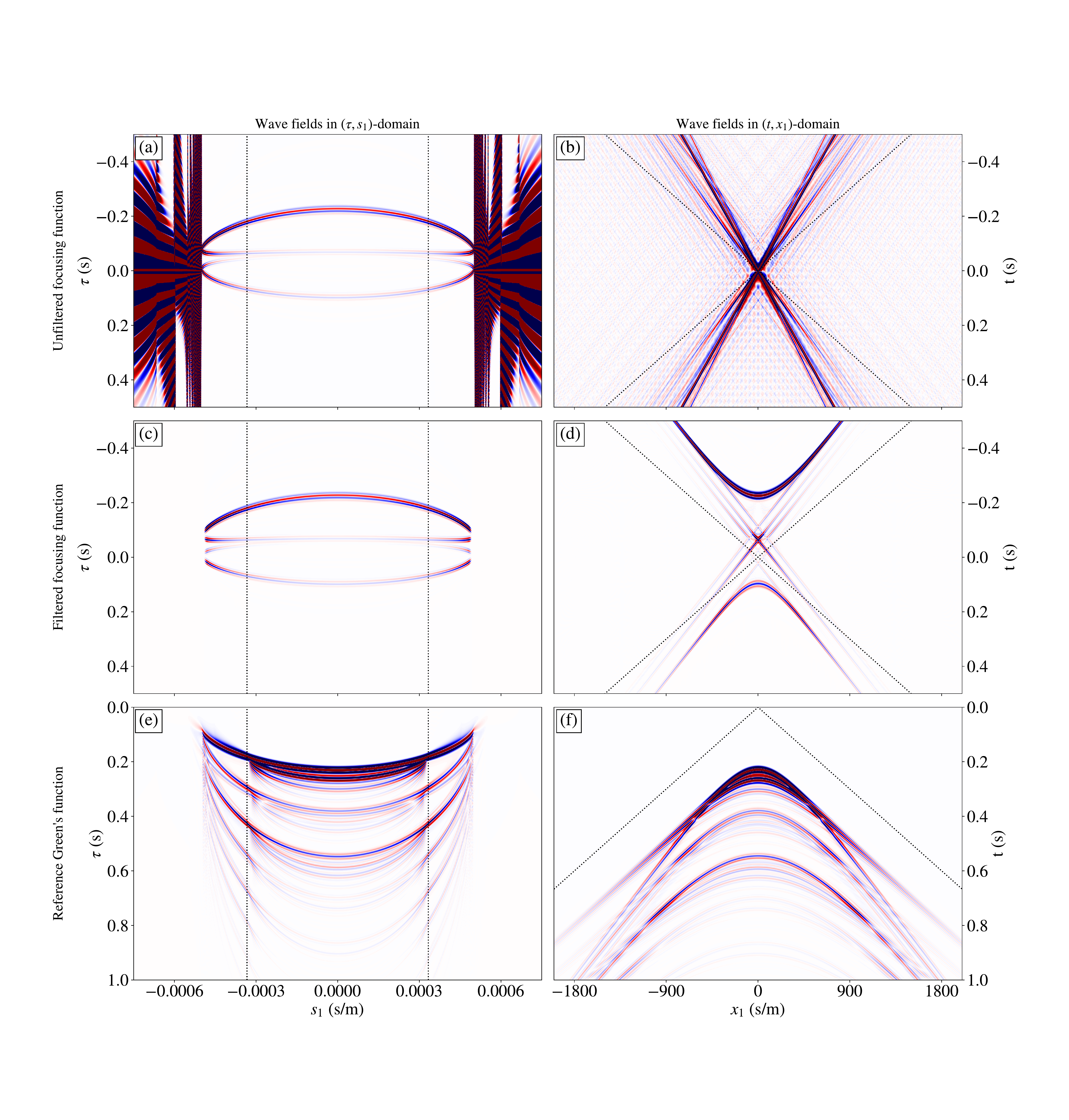}
\caption{wave fields in the subsurface for a focal position of (0,415)m shown in the $(\tau,s_1)-$domain in the left column and the $(t,x_1)-$domain in the right column. (a) Reference focusing function obtained from the full medium without filtering applied. (b) Transformed version of the unfiltered data in (a). (c) Same as (a), but with filtering for the high slowness values applied. (d) Transformed version of (c). (e) Modeled Green's function for various values of slowness and intercept time. (f) Modeled Green's function, obtained through the use of the \texttt{Salvus} modeling package. All wave fields have been convolved with a 50Hz Ricker wavelet. The dotted lines in the left column indicate the critical slowness values between propagating and evanescent waves at the focal depth. The dotted lines in the right column display the angles of the critical slowness value.}
\label{fig:refdat}
\end{figure}

The results in the previous sections are encouraging, so we consider whether this approach can be used to retrieve wave fields in the space-time domain. In order to do this, we use the workflow that we showed in the previous section and repeat it for many different values of slowness. The results are then transformed back to the space-time domain using an inverse Radon transform \cite{ravasi2020pylops}. An example is shown in \figref{fig:refdat}. The left column of this figure shows the focusing functions and Green's functions in the slowness-intercept time domain and the right column shows the same functions in the space-time domain. In \figref{fig:refdat} we show the reference focusing function that we obtain using \eqnsref{f1pT}-\eqref{fpcond} at a focal depth of 415m. The slowness value where the wave field transitions from propagating to evanescent is indicated by the vertical dotted black line. Note that for high absolute values of horizontal slowness, the wave field becomes very unstable. This is because the transmission response tends towards zero and the inversion in \eqnref{f1pT} blows up as a result. When this wave field is transformed to the space-time domain, the result is a mess, as can be seen in \figref{fig:refdat}b, where the dotted line indicates the angle of the slowness value where the wave field becomes evanescent. The result is dominated by strong artifacts. To mitigate these effects, a taper is applied at the edge of the zone of instability. This is shown in the second row. As can be seen, in both domains, the results are much cleaner. These are the wave fields that are used as a reference for the retrieved wave fields. In the third row, we show the reference Green's function. The Green's function in both domains were retrieved independently from each other. The reference in the slowness-intercept time domain was retrieved by a 1D modeling code, while the result in the space-time domain was obtained using the finite element modeling code \texttt{Salvus}, which was developed by Mondaic AG. Note that while for the focusing function there does not appear to be a clear border between the propagating and evanescent wave field in the left column, there is a clear difference between the propagating and evanescent wave field for the Green's function, where multiple events start converging.

%%%%%%%%%%%%%%%%%%%%%%%%%%%%%%%%%%%%%%%%%%%%%%%%%%%%%%%%%%%%%%%%%%%%%%%%%%%%%%%%%%%%%%%%%%%%%%%%%%%%%%%%%%%%%%%%%%%%%%%%%%%%%
%%%%%%%%%%%%%%%%%%%%%%%%%%%%%%%%%%%%%%%%%%%%%%%%%%%%%%%%%%%%%%%%%%%%%%%%%%%%%%%%%%%%%%%%%%%%%%%%%%%%%%%%%%%%%%%%%%%%%%%%%%%%%

\begin{figure}[!htpb]
\centering
\includegraphics[clip,trim={2cm 7cm 1cm 5cm},width=0.95\textwidth]{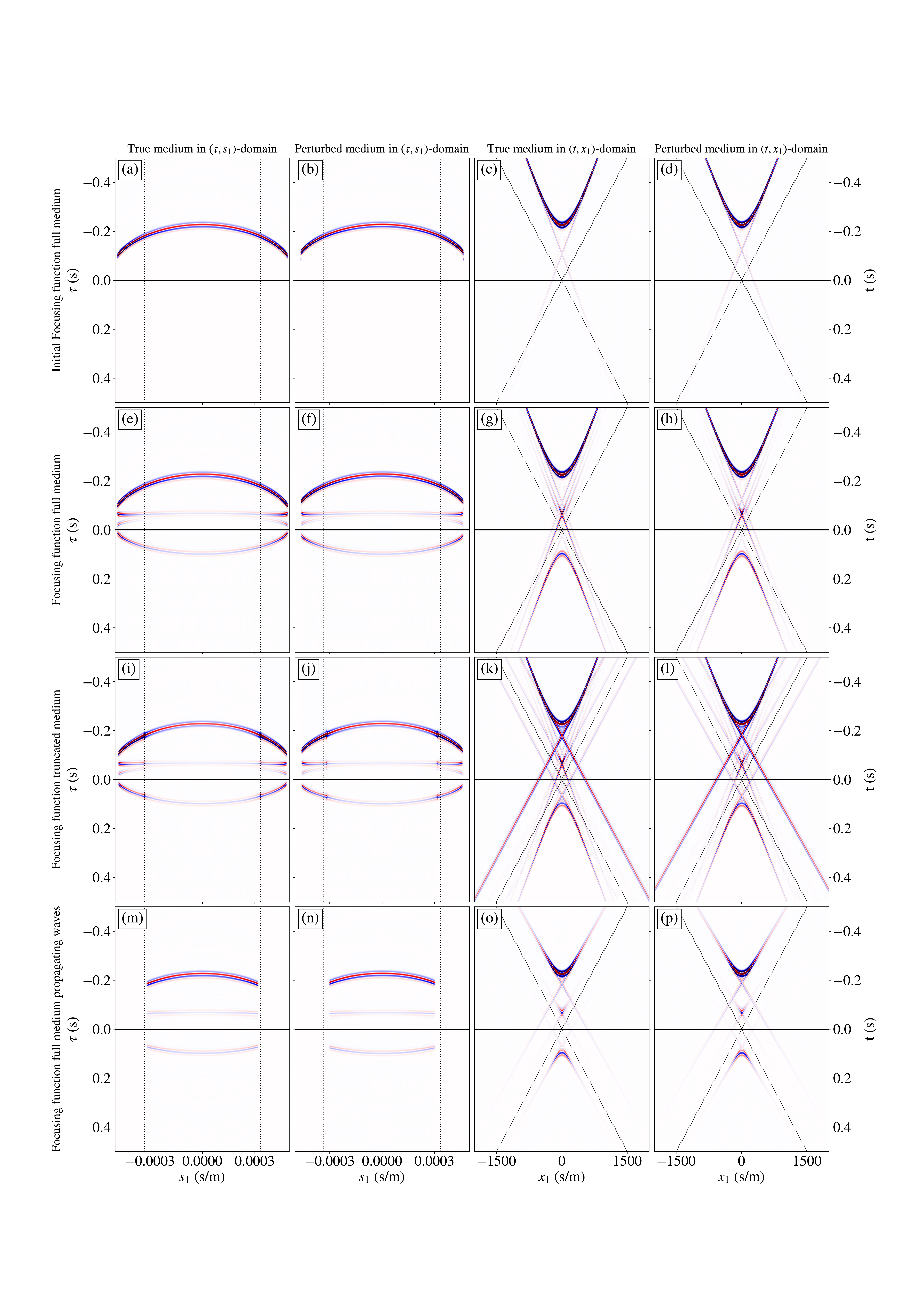}
\caption{Various versions of the focusing function. The time-zero axis has been indicated by the horizontal solid black line. The first row shows the estimation of the focusing function using the full medium. The second and third row show the result of the Marchenko method using the initial estimation from the full medium and the truncated medium, respectively. The fourth row shows the retrieved focusing function when only propagating waves are considered.
The first two columns show the data in the $(\tau,s_1)-$domain and the last two columns show the data after they have been transformed to the $(t,x_1)-$domain. The first and third column show the data that have been retrieved using the true medium parameters and the second and fourth column show the data that have been retrieved using the perturbed medium. All wave fields have been convolved with a 50Hz Ricker wavelet. The dotted lines in the first two columns indicate the critical slowness values between propagating and evanescent waves. The dotted lines in the final two columns display the angles of the critical slowness values.}
\label{fig:focptxt}
\end{figure}

In \figref{fig:focptxt}, we show the results of the Marchenko method in both domains. The first two columns show the wave fields in the slowness-intercept time domain and the final two columns show the same wave fields in the space-time domain. In the first and third column on the first row, we show the direct arrival that was retrieved from the full true medium and the second and fourth columns show the direct arrival that was obtained using the perturbed medium instead. The second row shows the retrieval of the Marchenko method using this direct arrival. Note that the results in both domains are very similar to each other, which demonstrates that the method is not overly sensitive to incorrect medium parameters. The effect of using the truncated medium instead of the full medium for the direct arrival is very significant however, which can be seen in the third row. In the slowness-intercept time domain, there is a very clear jump present between the propagating and evanescent wave field, without a smooth transition. This results in strong artifacts in the space-time domain, that have the same angle as the slowness value between the propagating and evanescent wave field. In this case, it is recommended to ignore the evanescent wave field entirely and just consider the propagating part of the wave field. In the fourth row, we show the result of the third row after the evanescent part of the wave field was filtered out. This is essentially the result that the classical Marchenko method is capable of retrieving. It can be seen in the space-time domain that the previous artifact has been greatly suppressed, however, the desired part of the wave field is also removed.

%%%%%%%%%%%%%%%%%%%%%%%%%%%%%%%%%%%%%%%%%%%%%%%%%%%%%%%%%%%%%%%%%%%%%%%%%%%%%%%%%%%%%%%%%%%%%%%%%%%%%%%%%%%%%%%%%%%%%%%%%%%%%
%%%%%%%%%%%%%%%%%%%%%%%%%%%%%%%%%%%%%%%%%%%%%%%%%%%%%%%%%%%%%%%%%%%%%%%%%%%%%%%%%%%%%%%%%%%%%%%%%%%%%%%%%%%%%%%%%%%%%%%%%%%%%

\begin{figure}[!htpb]
\centering
\includegraphics[clip,trim={2cm 7cm 1cm 5cm},width=0.95\textwidth]{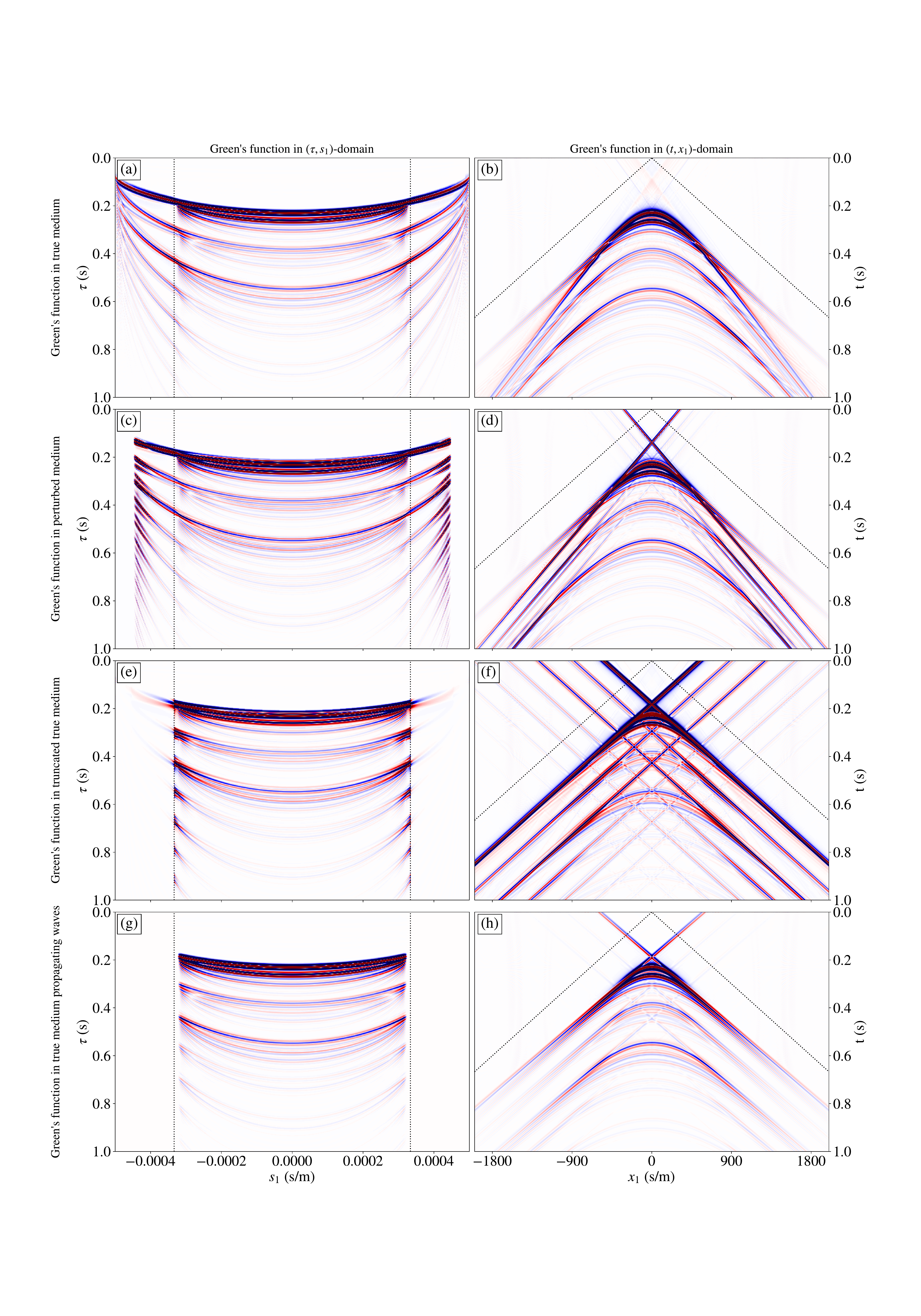}
\caption{Retrieved Green's functions using the focusing function in \figref{fig:focptxt}. The first and second column show the data in the $(\tau,s_1)-$domain and the $(t,x_1)-$domain, respectively. 
(a), (c), (e) and (g) show the Green's functions retrieved using the focusing function from \figref{fig:focptxt}(e), (f), (i) and (m), respectively.
All wave fields have been convolved with a 50Hz Ricker wavelet. The dotted lines in the left column indicate the critical slowness values between propagating and evanescent waves. The dotted lines in the right column display the angles of the critical slowness values.}
\label{fig:greenptxt}
\end{figure}

\figref{fig:greenptxt} shows the Green's functions that are retrieved using the various focusing functions from \figref{fig:focptxt} in the slowness-intercept time domain in the left column and the space-time domain in the right column. The first row in \figref{fig:greenptxt} shows the Green's function that was obtained from \eqnref{RFtp} using the focusing function from \figref{fig:focptxt}e. The result for both domains is accurate and the various events are retrieved. There are some weak artifacts present, caused by the inverse Radon transform. The second row shows the result obtained using the focusing function from \figref{fig:focptxt}f. The majority of the events are still retrieved in \figref{fig:greenptxt}c, however, for high values of absolute horizontal slowness, some issues arise. This is even more clear in the space-time domain, as can be seen from \figref{fig:greenptxt}d, where stronger artifacts are present. The majority of the wave field still appear to be retrieved. The retrieval using the focusing function from \figref{fig:focptxt}i is poor, even though the correct medium parameters were used. The evanescent wave field is not retrieved at all and the transition between the propagating and evanescent wave field is marked with strong artifacts. This results in a contaminated result in the space-time domain. Filtering out the evanescent part of the wave field removes most of these artifacts, as can be seen in the fourth row of \figref{fig:greenptxt}, however, once again, this also removes a large part of the desired wave field.

%%%%%%%%%%%%%%%%%%%%%%%%%%%%%%%%%%%%%%%%%%%%%%%%%%%%%%%%%%%%%%%%%%%%%%%%%%%%%%%%%%%%%%%%%%%%%%%%%%%%%%%%%%%%%%%%%%%%%%%%%%%%%
%%%%%%%%%%%%%%%%%%%%%%%%%%%%%%%%%%%%%%%%%%%%%%%%%%%%%%%%%%%%%%%%%%%%%%%%%%%%%%%%%%%%%%%%%%%%%%%%%%%%%%%%%%%%%%%%%%%%%%%%%%%%%

\begin{figure}[!htpb]
\centering
\includegraphics[clip,trim={4cm 7cm 4cm 8cm},width=0.95\textwidth]{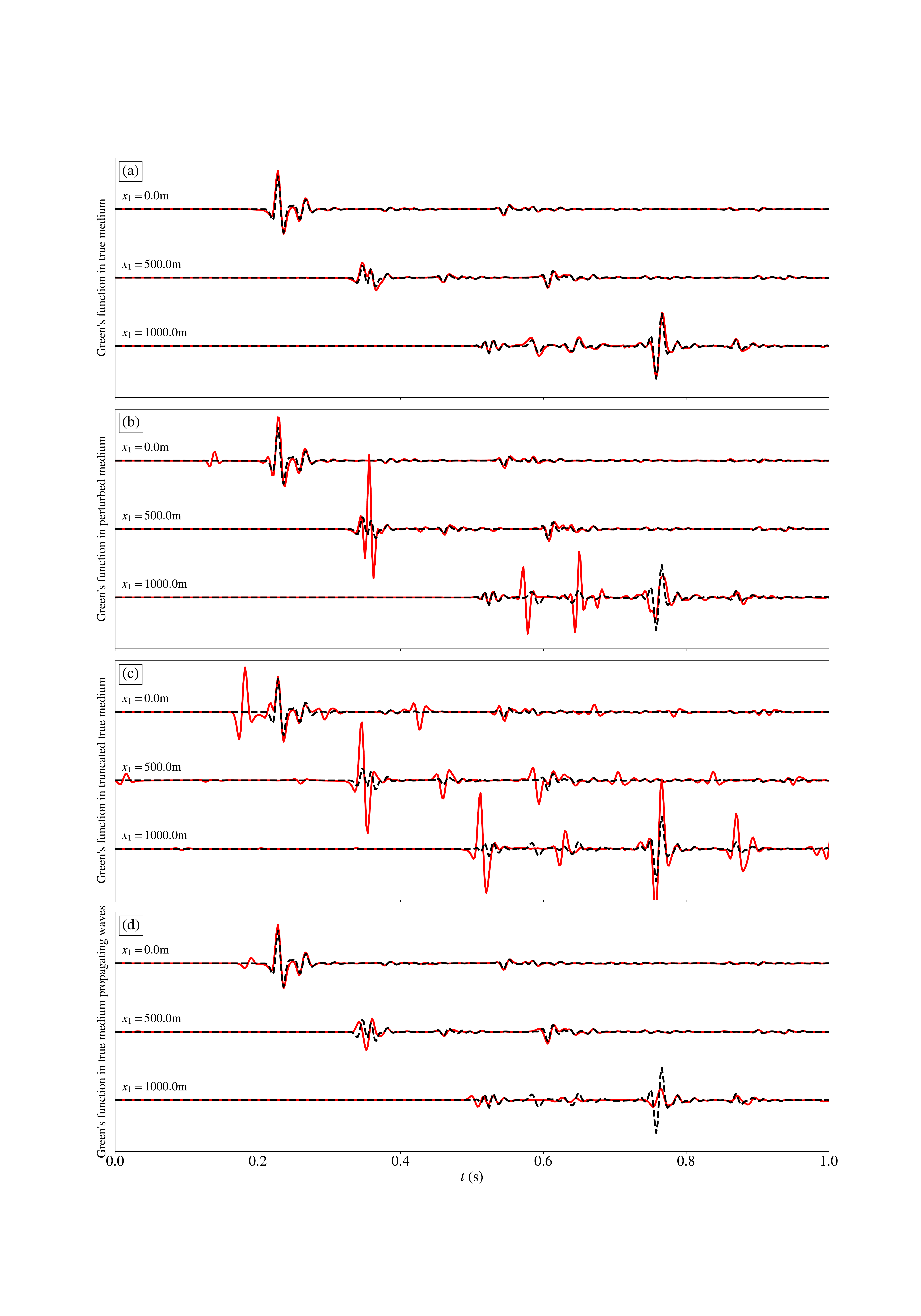}
\caption{Traces of the retrieved Green's functions in the $(t,x_1)-$domain, indicated in solid red, compared to traces at the same positions of the reference Green's function obtained with the \texttt{Salvus} modeling package, from \figref{fig:refdat}(f), shown as the dashed black lines. The traces shown are from (a) \figref{fig:greenptxt}(b), (b) \figref{fig:greenptxt}(d), (c) \figref{fig:greenptxt}(f) and (d) \figref{fig:greenptxt}(h). All wave fields have been convolved with a 50Hz Ricker wavelet. Each trace in a subfigure corresponds to a different offset.}
\label{fig:traces}
\end{figure}

To gauge the accuracy of the retrieved wave fields in the space-time domain more accurately, we compare traces of the results of the right column in \figref{fig:greenptxt} to the result of the modeled wave field in \figref{fig:refdat}f directly. We show these comparisons in \figref{fig:traces}, where the black dashed lines show the modeled result and the solid red lines in each row are the retrieved results that correspond to the rows of \figref{fig:greenptxt}. From \figref{fig:traces}, we can see that the result using the true direct arrival is very accurate for all offsets. When the direct arrival is estimated using the perturbed medium, the wave field is very accurate for low offsets, but the presence of artifacts increases for higher offsets. A large part of the wave field is still accurately retrieved. A specific slope filter could be used to mitigate these effects. When the truncated medium is used, the results are poor for all offsets, which is to be expected. When only propagating waves are considered, we can see that for small offsets, the results are accurate, but for higher offsets, certain information is missing.

%%%%%%%%%%%%%%%%%%%%%%%%%%%%%%%%%%%%%%%%%%%%%%%%%%%%%%%%%%%%%%%%%%%%%%%%%%%%%%%%%%%%%%%%%%%%%%%%%%%%%%%%%%%%%%%%%%%%%%%%%%%%%
%%%%%%%%%%%%%%%%%%%%%%%%%%%%%%%%%%%%%%%%%%%%%%%%%%%%%%%%%%%%%%%%%%%%%%%%%%%%%%%%%%%%%%%%%%%%%%%%%%%%%%%%%%%%%%%%%%%%%%%%%%%%%
\subsection*{Green's function retrieval}

\begin{figure}[!htpb]
\centering
\includegraphics[width=0.95\textwidth]{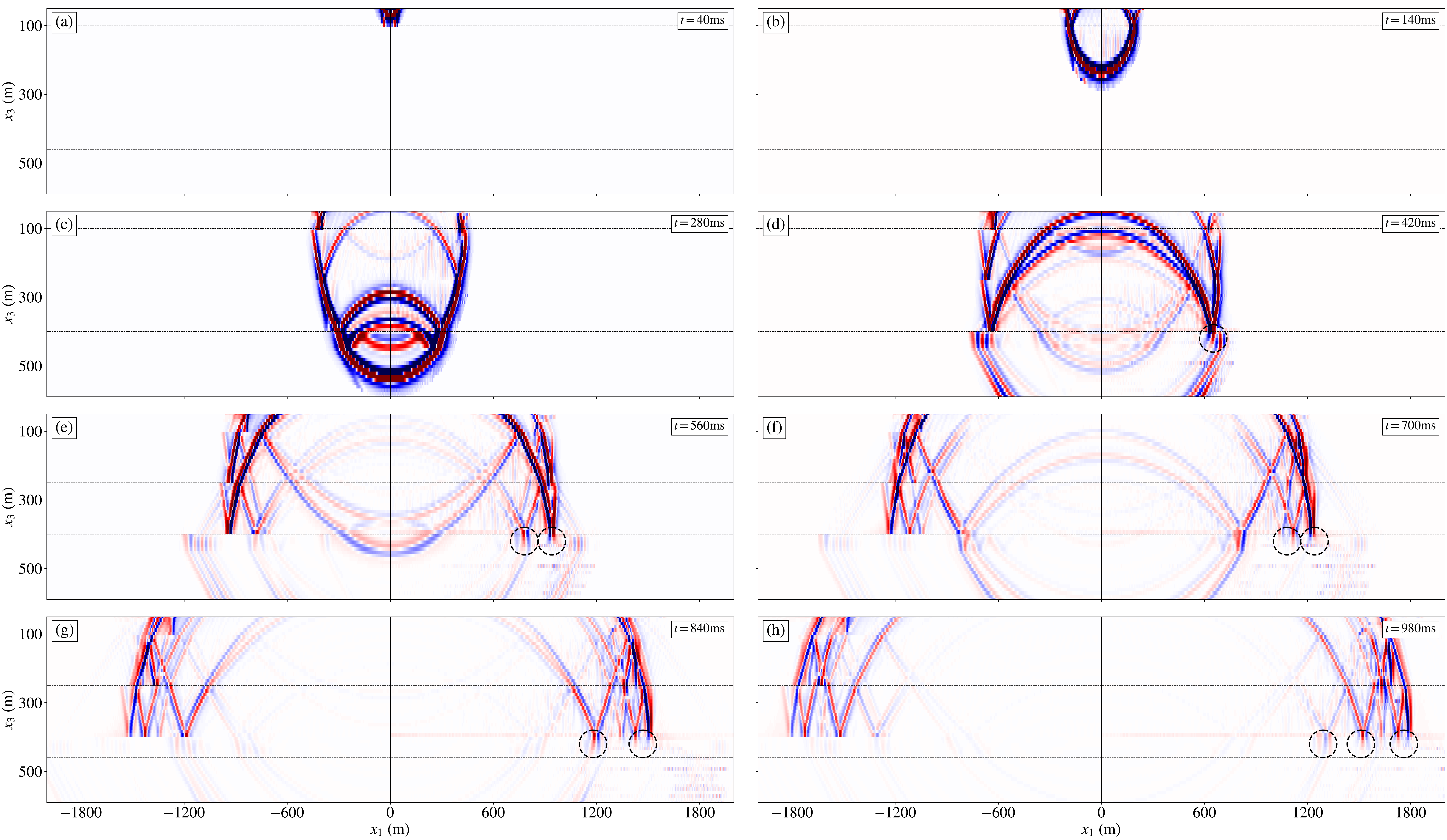}
\caption{Green's functions retrieved in the $t-\bx$-domain, where the source is located at the surface of the Earth. The left side of each panel shows the wave field that was retrieved when only propagating waves were considered. The right side of the panels show the same results when both the propagating and evanescent part of the wave field were retrieved. The solid black vertical line indicates the border between these results. In the final four panels, the evanescent part of the wave field that was retrieved is indicated by the black dashed cirecles. The horizontal dashed lines indicate the locations of boundaries between the layers. All wave fields have been convolved with a 50Hz Ricker wavelet.}
\label{fig:Gret}
\end{figure}

The previous section demonstrated that the Green's function that is retrieved when the propagating and evanescent wave field are both considered is more accurate. To demonstrate how this wave field travels through the subsurface, we retrieve the Greens' function for various depths, while keeping the source at the surface of the Earth. Each procedure is done in the exact same way. The retrieved wave field is shown in \figref{fig:Gret}, where each panel shows a snapshot of the subsurface at a different time. To show what the retrieval of the evanescent part of the wave field adds, we have split each panel into two parts. The left part of each panel shows the Green's function when only the propagating part of the wave field is retrieved and the right part of the panel shows the wave field when both the propagating and evanescent parts of the wave field are retrieved. Because the medium is laterally invariant, the wave field can be mirrored over the center. The vertical solid black line indicates the border between these parts. In the upper three layers of the medium, the difference between the retrieved wave fields is very minor, but in the fourth layer, we can see the difference. At 420ms, we can see that there is an additional event at the right side, which we indicated with a black dashed circle, that is not present on the left side. When we study the later times, we can see that more of these events appear. These are clearly refracted waves, built up from the evanescent wave field, as they travel horizontally and the amplitude decays with depth. We can also see that the frequency content of the wave field is affected, depending on the depth. What is particularly important to note is that not only the direct arrival of the Green's function has this evanescent part of the wave field, but the multiples of the wave field as well. This clearly shows that the Marchenko method can retrieve multiply refracted waves.

%%%%%%%%%%%%%%%%%%%%%%%%%%%%%%%%%%%%%%%%%%%%%%%%%%%%%%%%%%%%%%%%%%%%%%%%%%%%%%%%%%%%%%%%%%%%%%%%%%%%%%%%%%%%%%%%%%%%%%%%%%%%%
%%%%%%%%%%%%%%%%%%%%%%%%%%%%%%%%%%%%%%%%%%%%%%%%%%%%%%%%%%%%%%%%%%%%%%%%%%%%%%%%%%%%%%%%%%%%%%%%%%%%%%%%%%%%%%%%%%%%%%%%%%%%%

\newpage
\section*{Conclusion and Outlook}
We have shown that the Marchenko method can retrieve the evanescent part of the wave field, not only for the direct reflections, but also the multiply reflected wave field. The evanescent part of the multiply reflected wave field constitutes multiply refracted waves. In order to achieve this, it is critical that the evanescent part of the wave field is represented in the direct arrival that is used in the iterative Marchenko scheme. Future work should aim at retrieving the full wave field including the propagating and evanescent waves when the wave field redatumed at both the source and receiver side.

\section*{acknowledgments}
We acknowledge funding from the European Research Council (ERC) under the European Union’s Horizon 2020 research and innovation programme (grant agreement no.: 742703). This work has received support from the Swiss National Supercomputing Centre CSCS and Mondaic AG.

\newpage
% References
\bibliography{references}

\end{document}